\documentclass[preprint,aps,prd,superscriptaddress,nofootinbib,tightenlines]{revtex4}

\usepackage{amsfonts}
\usepackage{multirow}
\usepackage{mathrsfs}
\usepackage{graphicx}
\usepackage{amsmath}
\usepackage{amssymb}
\usepackage{bm}
\usepackage{bbm}

\usepackage{color}
\usepackage{hyperref}
\usepackage{ulem}
\usepackage{subfig}
\usepackage{caption}
\usepackage{slashed}
\usepackage{mathtools}
\usepackage{float}

\def\OMIT#1{}

\newcommand{\beq}{\begin{equation}}
\newcommand{\eeq}{\end{equation}}
\newcommand{\bqa}{\begin{eqnarray}}
\newcommand{\eqa}{\end{eqnarray}}

\newcommand{\bseq}{\begin{subequations}}
\newcommand{\eseq}{\end{subequations}}

\newcommand{\dd}{{\mathrm{d}}}
\newcommand{\ket}[1]{\left\lvert{#1}\right\rangle}

\newcommand{\expval}[1]{\left\langle{#1}\right\rangle}
\newcommand{\abs}[1]{\left\lvert{#1}\right\rvert}

\newcommand{\triplet}{\ensuremath{\mathbf{\bar{3}}\otimes\mathbf{3}}}
\newcommand{\sextet}{\ensuremath{\mathbf{6}\otimes\mathbf{\bar{6}}}}

\begin{document}


\title{\mbox{}\\[14pt]
Inclusive production of fully-charmed tetraquarks at LHC}

\author{Feng Feng~\footnote{F.Feng@outlook.com}}
\affiliation{China University of Mining and Technology, Beijing 100083, China\vspace{0.2 cm}}
\affiliation{Institute of High Energy Physics, Chinese Academy of
  Sciences, Beijing 100049, China\vspace{0.2 cm}}

\author{Yingsheng Huang~\footnote{yingsheng.huang@northwestern.edu}}
\affiliation{High Energy Physics Division, Argonne National Laboratory, Argonne, IL 60439, USA}
\affiliation{Department of Physics \& Astronomy,
  Northwestern University, Evanston, IL 60208, USA}

\author{Yu Jia~\footnote{jiay@ihep.ac.cn}}
\affiliation{Institute of High Energy Physics, Chinese Academy of
  Sciences, Beijing 100049, China\vspace{0.2 cm}}
\affiliation{School of Physics, University of Chinese Academy of Sciences,
  Beijing 100049, China\vspace{0.2 cm}}

\author{Wen-Long Sang~\footnote{wlsang@swu.edu.cn}}
\affiliation{School of Physical Science and Technology, Southwest University, Chongqing 400700, P.R. China}

\author{De-Shan Yang\footnote{yangds@ucas.ac.cn}}
\affiliation{School of Physics, University of Chinese Academy of Sciences,
  Beijing 100049, China\vspace{0.2 cm}}
\affiliation{Institute of High Energy Physics, Chinese Academy of
  Sciences, Beijing 100049, China\vspace{0.2 cm}}

\author{Jia-Yue Zhang~\footnote{zhangjiayue@ihep.ac.cn}}
\affiliation{Institute of High Energy Physics, Chinese Academy of Sciences, Beijing 100049, China\vspace{0.2 cm}}
\affiliation{School of Physics, University of Chinese Academy of Sciences,
  Beijing 100049, China\vspace{0.2 cm}}

\date{\today}

\begin{abstract}
  The $X(6900)$ resonance, originally discovered by the \texttt{LHCb} collaboration and later confirmed by both \texttt{ATLAS} and \texttt{CMS} experiments, has sparked
  broad interests in the fully-charmed tetraquark states. Relative to the mass spectra and decay properties of fully-heavy tetraquarks,
  our knowledge on their production mechanism is still rather limited. In this work we investigate the inclusive production of fully-charmed $S$-wave
  tetraquarks at \texttt{LHC} within the nonrelativistic QCD (NRQCD) factorization framework. The partonic cross sections are computed at lowest order in $\alpha_s$ and velocity,
  while the long-distance NRQCD matrix elements are estimated from phenomenological potential models.
  We predict the differential $p_T$ spectra of various fully-charmed $S$-wave tetraquarks at the \texttt{LHC}, and compare with the results predicted from the
  fragmentation mechanism at large $p_T$ end.
\end{abstract}

\maketitle

\paragraph{\color{blue} Introduction.}
In $2020$, the \texttt{LHCb} collaboration reported the unexpected discovery of a new resonance, dubbed $X(6900)$, in the di-$J/\psi$ invariant mass spectrum~\cite{Aaij:2020fnh}.
Later in 2022, both the \texttt{ATLAS} and \texttt{CMS} collaborations~\cite{ATLAS:2022hhx,CMS:2022yhl} confirmed the existence of this new particle.
$X(6900)$ is widely believed to be a viable candidate for the fully-charmed compact tetraquark state~\cite{2204.02649},
though other possibilities have also been explored~\cite{Wang:2020wrp,Dong:2020nwy,Guo:2020pvt,Gong:2020bmg, 2011.00978, 2104.08589}.
Since the charm quark is too heavy to be readily knocked out of the vacuum, the dynamic feature of the fully-charm tetraquark (hereafter denoted by $T_{4c}$) is
dominated by its leading Fock component $\ket{cc\bar c\bar c}$, thus free from the contamination by the light constitutes.
Analogous to the fact that heavy quarkonia are the simplest hadrons, the fully-charmed tetraquarks are the simplest exotic hadrons from theoretical perspective.

Long before the discovery of the $X(6900)$, the existence of possible fully-heavy tetraquark states have been explored since 1970s~\cite{Iwasaki:1976cn,Chao:1980dv,Ader:1981db}.
The mass spectra and decay properties of fully-heavy tetraquarks have been investigated from various phenomenological models, including quark potential models~\cite{Becchi:2020uvq,Lu:2020cns,liu:2020eha,Karliner:2020dta,Zhao:2020nwy,Zhao:2020cfi,Giron:2020wpx,Ke:2021iyh,Gordillo:2020sgc,Yang:2020rih,Jin:2020jfc,Mutuk:2022nkw,Wang:2022yes}
and QCD sum rules~\cite{Chen:2020xwe,Wang:2020ols,Yang:2020wkh,Wan:2020fsk,Zhang:2020xtb}.
On the other hand, the study of the production mechanism of fully-heavy tetraquarks is relatively sparse, which is mainly based on color evaporation model and duality relations~\cite{Karliner:2016zzc,Berezhnoy:2011xy,Berezhnoy:2011xn,Becchi:2020mjz,Becchi:2020uvq,Maciula:2020wri,Carvalho:2015nqf,Goncalves:2021ytq}.
Inspired by the unexpected discovery of the $X(6900)$, recently several groups have attempted to investigate the $T_{4c}$ production in the context of
model-independent NRQCD factorization framework~\cite{Ma:2020kwb,Feng:2020riv,Feng:2020qee,Huang:2021vtb,Zhu:2020xni}.
Ma and Zhang studied the inclusive production of $T_{4c}$ at \texttt{LHC} and conducted a numerical study of
the dependence of the ratio $\sigma(2^{++})/\sigma(0^{++})$ on $p_T$~\cite{Ma:2020kwb}.
Zhu computed the $gg\to T_{4c}$ channel and predicted the low-$p_T$ spectrum of the $T_{4c}$ at \texttt{LHC} utilizing Collins-Soper-Sterman resummation~\cite{Zhu:2020xni}.
Feng {\it et al.} explicitly introduced the NRQCD operators relevant to $S$-wave $T_{4c}$ production, and derived the approximate
relation between the long-distance NRQCD matrix elements and the tetraquark wave functions at the origin~\cite{Feng:2020riv}.
Feng {\it et al.} have applied the NRQCD factorization approach to predict the $T_{4c}$ hadroproduction at large $p_T$ via fragmentation mechanism~\cite{Feng:2020riv}, as well as
inclusive and exclusive production of $T_{4c}$ at $B$ factories~\cite{Feng:2020qee,Huang:2021vtb}.

The goal of this work is to apply the NRQCD factorization approach elaborated in ~\cite{Feng:2020riv} to investigate
the $p_T$ spectrum of the $S$-wave $T_{4c}$ at $\texttt{LHC}$. We focus on the $gg\to T_{4c}+g$ channels and compute the
short-distance coefficients at lowest order in $\alpha_s$ and $v$.  We appeal to phenomenological potential models to estimate
the nonperturbative NRQCD matrix elements. The numerical studies indicate that there are bright prospects to measure the $p_T$
spectrum of $T_{4c}$.
Since the bulk of cross sections come from the small-$p_T$ regime,
where the fragmentation mechanism fails to be applicable, we hope that
our fixed-order NRQCD prediction provides more useful guidance for future experimental measurements of the $T_{4c}$ spectrum.

\paragraph{\color{blue} NRQCD Factorization formula for $T_{4c}$ hadroproduction \label{sec:factorization}}
According to QCD factorization theorem, the inclusive production rate of the fully-charmed tetraquark $T_{4c}$ in hadronic collisions can be expressed as
\begin{align}
\dd\sigma(p p & \rightarrow T_{4c}+X)=\sum_{i, j=q, g} \int_0^1 d x_1 d x_2 f_{i / p}\left(x_1, \mu_F\right)
f_{j / p}\left(x_2, \mu_F\right) \dd\hat{\sigma}_{ij\to T_{4c}+X}(x_1x_2s, \mu_F),
\label{QCD:factorization:theorem}
\end{align}
where $f_{i/p}(x, \mu_F)$ denotes the parton distribution function (PDF) of the parton $i$ inside the proton, and $\mu_F$
represents the factorization scale.
$\hat{\sigma}_{ij\to T_{4c}+X}(x_1x_2s,\mu_F)$ is the partonic cross section for the $ij\to T_{4c}+X$ channel.
If we are interested in the $T_{4c}$ production with not overly large $p_T$,
since the gluon density is much more dominant than quark density at small $x$,
it suffices to only consider the gluon-gluon fusion and neglect the $q\bar{q}$ channel.

The partonic cross section $\hat{\sigma}_{ij\to T_{4c}+X}(x_1x_2s,\mu_R,\mu_F)$ in \eqref{QCD:factorization:theorem} still
encapsulates the non-perturbative effects about the formation of the $T_{4c}$.
Since four charm quarks have to be created in relatively short distance to have a non-negligible chance to form $T_{4c}$,
owing to asymptotic freedom, one anticipates that NRQCD factorization can be invoked to further factorize the partonic
cross section $\hat{\sigma}_{T_{4c}+X}$ into the product of the perturbatively calculable short-distance coefficients (SDCs)
and the nonperturbative long-distance matrix elements (LDMEs):
\begin{align}
  \dfrac{\dd\hat{\sigma}_{T_{4c}+X}}{\dd\hat{t}}= \sum_n {F_n(\hat{s},\hat{t})\over {m_c^{14}} } (2M_{T_{4c}})\expval{O_n^{T_{4c}}},
\label{eq:NRQCD_factorization}
\end{align}
where $\dd\hat{\sigma}_{T_{4c}+X}$ signifies the partonic cross section for $gg\to T_{4c}+X$.
$F_n$ denotes the SDC associated with different color configuration $n$,  and $\expval{O_n^{T_{4c}}}$ denotes the vacuum matrix elements of various
NRQCD production operators. $\hat{s}$ and $\hat{t}$ are the usual partonic Mandelstam variables.
The factor $2M_{T_{4c}}$ is inserted to compensate for the fact that the $T_{4c}$ state is nonrelativistically normalized in the LDMEs.

The main concern of this work is about the $S$-wave fully-charmed tetraquarks, which may carry the $J^{PC}$ quantum number of
$0^{++}$, $1^{+-}$ and $2^{++}$. It is convenient to adopt the diquark basis to specify the color configuration.
In this context, the color-singlet tetraquark is decomposed into either $\triplet$ or $\sextet$ diquark-antidiquark clusters.
The former case corresponds to the spin-1 diquark, while the latter corresponds to the spin-0 diquark.

Specifically speaking, at the lowest order in velocity expansion, Eq.~\eqref{eq:NRQCD_factorization} takes the following form:
\beq
{\dd\hat{\sigma} ( T_{4c}^{(J)}+X) \over \dd\hat{t}}
 =  \frac{2M_{T_{4c}}}{m_c^{14}} \left[
 F_{3,3}^{(J)} \expval{{O}^{(J)}_{3,3}}+ 2 F_{3,6}^{(J)} \expval{{O}^{(J)}_{3,6}} +F_{6,6}^{(J)}\expval{{O}^{(J)}_{6,6}}\right],
\label{eq:factorized:cross:section}
\eeq
with $J=0,1,2$.
${O}^{(J)}_{\mathrm{color}}$ denote the NRQCD production operators with different color configuration,
which were introduced in \cite{Feng:2020riv,Feng:2020qee,Huang:2021vtb}:
\begin{subequations}
  \begin{align}
     & {O}^{(J)}_{3,3} = \mathcal{O}^{(J)}_{\triplet}\sum_X|T_{4c}^J+X\rangle\langle T_{4c}^J+X|\mathcal{O}^{(J)\dagger}_{\triplet},
\\
& {O}^{(0)}_{6,6} = \mathcal{O}^{(0)}_{\sextet}\sum_X|T_{4c}^0+X\rangle\langle T_{4c}^0+X|\mathcal{O}^{(0)\dagger}_{\sextet},
\\
& {O}^{(0)}_{3,6} = \mathcal{O}^{(0)}_{\triplet}\sum_X|T_{4c}^0+X\rangle\langle T_{4c}^0+X|\mathcal{O}^{(0)\dagger}_{\sextet},
\end{align}
\end{subequations}
with the quartic NRQCD operators $\mathcal{O}^{(J)}_{\triplet}$ and $\mathcal{O}^{(0)}_{\sextet}$ defined by
\begin{subequations}
\begin{align}
     & \mathcal{O}^{(0)}_{\triplet}=-\frac{1}{\sqrt{3}}[\psi_a^T(i\sigma^2)\sigma^i\psi_b] [\chi_c^{\dagger}\sigma^i (i\sigma^2)\chi_d^*]\;
    \mathcal{C}^{ab;cd}_{\triplet},
    \\
     & \mathcal{O}^{i;(1)}_{\triplet}= -{\frac{i}{\sqrt{2}}} \left[\psi_a^T (i \sigma^2)\sigma^j\psi_b\right]\left[\chi_c^\dagger\sigma^k (i \sigma^2)\chi_d^*\right]\,\epsilon^{ijk}\;{\mathcal C}^{ab;cd}_{\triplet} ,
    \\
     & \mathcal{O}^{ij;(2)}_{\triplet} =[\psi_a^T(i\sigma^2)\sigma^m\psi_b] [\chi_c^{\dagger}\sigma^n(i\sigma^2)\chi_d^*]\;{\Gamma^{ij;mn}}
    \;\mathcal{C}^{ab;cd}_{\triplet},
    \\
     & {\mathcal{O}^{(0)}_{\sextet} =
          [\psi_a^T(i\sigma^2)\psi_b] [\chi_c^{\dagger}(i\sigma^2)\chi_d^*]\;
        \mathcal{C}^{ab;cd}_{\sextet}}.
  \end{align}
\label{NRQCD:composite:operators}
\end{subequations}
The color indices $a,b,c,d$ run from $1$ to $3$, and the Cartesian indices $i,j,k$ run from $1$ to $3$.
The rank-$4$ Lorentz tensor is given by {$\Gamma^{kl;mn}\equiv \frac{1}{2}(\delta^{k m} \delta^{l n}+\delta^{k n} \delta^{l m}-\frac{2}{3} \delta^{k l} \delta^{mn})$},
and the rank-$4$ color tensors $\mathcal{C}$ are defined as
\begin{subequations}
\begin{align}
& {\mathcal{C}^{ab;cd}_{\triplet}\equiv \frac{1}{(\sqrt{2})^2} \epsilon^{abm}\epsilon^{cdn}\frac{\delta^{mn}}{\sqrt{N_c}}=\frac{1}{2\sqrt{3}}(\delta^{ac}\delta^{bd}-\delta^{ad}\delta^{bc})},
\\
     & \mathcal{C}^{ab;cd}_{\sextet}
    \equiv \frac{1}{2\sqrt{6}}(\delta^{ac}\delta^{bd}+\delta^{ad}\delta^{bc}).
\end{align}
\label{color:tensor}
\end{subequations}

\paragraph{\color{blue}Determination of short-distance coefficients \label{sec:SDC}}

We employ the perturbative matching procedure to calculate various SDCs associated with different color channel in Eq.~(\ref{eq:factorized:cross:section}).
Since the SDCs are insensitive to the long-distance dynamics, one may replace the physical tetraquark states with free four-quark states $\ket{[cc][\bar c\bar c]}$ in
(\ref{eq:factorized:cross:section}).  It is then straightforward to calculate both sides 
in perturbative QCD  and perturbative NRQCD to solve for the SDCs.

\begin{figure}
\centering
\includegraphics{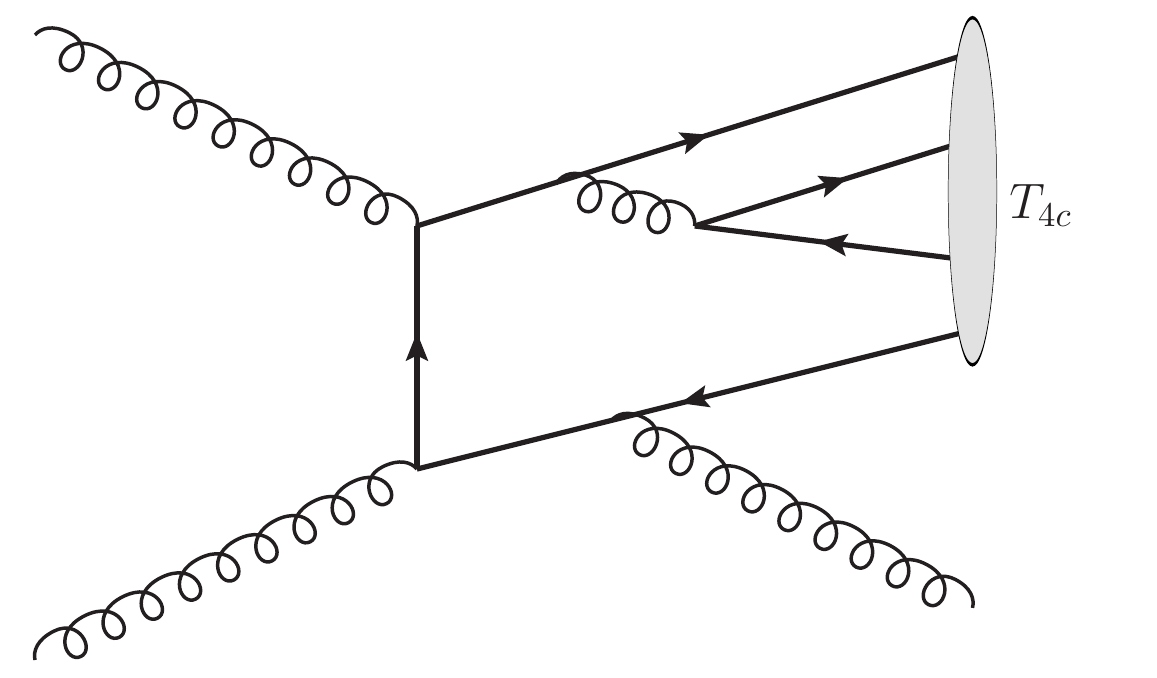}
\caption{One typical Feynman diagram for $gg\to T_{4c}+g$.}
\label{fig:feynman-diagram}
\end{figure}

We have normalized the NRQCD operators such that all vaccum-to-tetraquark matrix elements of the NRQCD composite operators are equal to $4$ (up to a factor of the polarization vector).
For the QCD part, we compute the amplitude of $gg\to \ket{[cc][\bar c\bar c]}+g$, employing the covariant color/Lorentz projector method to project out the desired amplitude
where the fictitious tetraquark states carry appropriate color/spin/orbital quantum numbers~\cite{Feng:2020riv}.
We work in Feynman gauge, and take $P^2=M^2_{T_{4c}}\approx 16m_c^2$ for simplicity. We employ our self-written program \texttt{HepLib}~\cite{Feng:2021kha} and \texttt{FeynArts}/\texttt{FeynCalc}~\cite{Hahn:2000kx,Shtabovenko:2016sxi} to generate the Feynman diagrams and square the amplitudes as well as sum over polarizations.
There are 642 Feynman diagrams in total, with one typical Feynman diagram shown in Fig.~\ref{fig:feynman-diagram}.
To avoid the occurrence of the ghost contribution, we only sum over two transverse polarizations for the incoming and outgoing gluons upon squaring the amplitude.
We have explicitly verified gauge invariance of the squared amplitude.

The complete expressions of the SDCs are too lengthy to be reproduced in the text.
For the convenience of the readers, we have attached those expressions in
an auxiliary file. 
Here we are contented with producing the asymptotic behaviors of various SDCs in the large $p_T$ limit:
\begingroup
\begin{subequations}
  \begin{align}
    &F_{3,3}^{0^{++}}= \frac{2209 \pi ^4 m_c^6 \alpha _s^5 \left(\hat{s} \hat{t}+\hat{s}^2+\hat{t}^2\right)^4}{15552 \hat{s}^5 (-\hat{t})^3 \left(\hat{s}+\hat{t}\right)^3}+\mathcal{O}\left(\frac{m_c^7}{p_T^7}\right),
\\
& F_{3,6}^{0^{++}}=\sqrt{6}F_{3,3}^{0^{++}}+\mathcal{O}\left(\frac{m_c^7}{p_T^7}\right),
\\
& F_{6,6}^{0^{++}}=\frac{2}{3}F_{3,3}^{0^{++}}+\mathcal{O}\left(\frac{m_c^7}{p_T^7}\right),
\\
& F_{3,3}^{1^{+-}}=\frac{60025 \pi ^4 m_c^8 \alpha _s^5 \left(\hat{s} \hat{t}+\hat{s}^2+\hat{t}^2\right)^2}{34992 \hat{s}^4 \hat{t}^2 \left(\hat{s}+\hat{t}\right)^2}+\mathcal{O}\left(\frac{m_c^9}{p_T^9}\right),
\\
& F_{3,3}^{2^{++}}= \frac{17617  \pi ^4 m_c^6 \alpha _s^5 \left(\hat{s} \hat{t}+\hat{s}^2+\hat{t}^2\right)^4}{38880 \hat{s}^5  (-\hat{t})^3 \left(\hat{s}+\hat{t}\right)^3}+\mathcal{O}\left(\frac{m_c^7}{p_T^7}\right),
\end{align}
\label{eq:SDC:large:pT}
\end{subequations}
\endgroup
with $\hat{s}$ and $\hat{t}$ scaling as $\mathcal{O}(p_T^{2})$.
Notice that all the five SDCs are positive.
We also observe that, the partonic cross sections for $C$-even tetraquarks  scale as $p_T^{-6}$, while those for the $C$-odd state $1^{+-}$
scale as $\mathcal{O}(p_T^{-8})$.  Thus the production rate of $C$-odd tetraquark is severely suppressed with respect to the $C$-even ones
due to the extra power of $p_T^{-2}$.  At large $p_T$, the predicted LO cross sections for the $C$-even states
receive an extra suppression factor of $p_T^{-2}$ with respect to the production rates predicted from the
fragmentation mechanism~\cite{Feng:2020riv}, which scale as $p_T^{-4}$.

\paragraph{\color{blue}Phenomenology of $T_{4c}$ production at the {LHC} \label{sec:phenomenology}}
A key ingredient in making concrete predictions is the LDMEs that enter the NRQCD factorization formula 
(\ref{eq:factorized:cross:section}). 
These nonperturbative matrix elements can in principle be calculated by the lattice NRQCD in the future. 
As a workaround, we appeal to phenomenological approaches to roughly estimate the values of these LDMEs.

After applying the vacuum saturation approximation, one may express the NRQCD LDMEs in terms of the wave functions at 
the origin of a tetraquark~\cite{Feng:2020riv,Feng:2020qee,Huang:2021vtb}:
\begin{subequations}
  \begin{align}
     & \expval{{O}_{C_1,C_2}^{(0)}}\approx 16\,{\psi_{C_1}(\mathbf{0})\psi_{C_2}^*(\mathbf{0})},
\\
& \expval{{O}_{C_1,C_2}^{(1)}}\approx 48\,{\psi_{C_1}(\mathbf{0})\psi_{C_2}^*(\mathbf{0})},
\\
& \expval{{O}_{C_1,C_2}^{(2)}}\approx 80\,{\psi_{C_1}(\mathbf{0})\psi_{C_2}^*(\mathbf{0})},
\end{align}
\label{LDME2wf}
\end{subequations}
where $\psi(\mathbf{0})$ denotes the four-body Schr\"odinger wave function at the origin. The color structure labels $C_1$ and $C_2$ can be either $3$ or $6$, 
representing the $\triplet$ and the $\sextet$ diquark-antidiquark configuration. 
In this work, we adopt two phenomenological potential models to estimate the wave functions at the origin~\cite{Lu:2020cns,liu:2020eha}.  
Both models assume Cornell-type spin-independent potential and incorporate some pieces of spin-dependent potentials, and numerically solve the four-body Schr\"odinger equation
using Gaussian basis.
As a slight difference, Model I is based on nonrelativistic quark potential model, while 
Model II utilizes the relativistic kinetic term. 
We enumerate the values of the predicted NRQCD LDMEs from both models in Table~\ref{tab:LDME}.

\begingroup
\setlength{\tabcolsep}{6pt} 
\renewcommand{\arraystretch}{2}
\begin{table}[H]
\centering
\begin{tabular}{cccc}
\hline
\hline
&\multicolumn{1}{c}{\textbf{LDME}} & \textbf{Model I}~\cite{Lu:2020cns} & \textbf{Model II}~\cite{liu:2020eha} \\
\hline
\multirow{3}{*}{$0^{++}$} & $\expval{{O}_{3,3}^{(0)}}[\mathrm{GeV}^9]$ & $0.0347$ & $0.0187$ \\
\cline{2-4}
&$\expval{{O}_{3,6}^{(0)}}[\mathrm{GeV}^9]$ & $0.0211$ & $-0.0161$ \\
\cline{2-4}
&$\expval{{O}_{6,6}^{(0)}}[\mathrm{GeV}^9]$ & $0.0128$ & $0.0139$ \\
\hline
$1^{+-}$&$\expval{{O}_{3,3}^{(1)}}[\mathrm{GeV}^9]$ & $0.0780$ & $0.0480$ \\
\hline
$2^{++}$&$\expval{{O}_{3,3}^{(2)}}[\mathrm{GeV}^9]$ & $0.072$ & $0.0628$ \\
\hline
\hline
\end{tabular}
\caption{Numerical values of the LDMEs estimated from Model I and Model II.}
\label{tab:LDME}
\end{table}
\endgroup

We then apply \eqref{eq:factorized:cross:section} in conjunction with \eqref{QCD:factorization:theorem} to predict 
the $p_T$ spectrum of the $T_{4c}$ in $pp$ collision at $\sqrt{s}=13\ \mathrm{TeV}$.  
We set the charm quark mass $m_c=1.5\ \mathrm{GeV}$ and use $\alpha_s(M_Z)=0.1180$~\cite{Dulat:2015mca}.
We adopt the \texttt{CT14lo} PDF set~\cite{Dulat:2015mca}, and impose a rapidity cut $\abs{y}\leq 5$.  
We choose the factorization scale $\mu_F=m_T$, with the transverse mass $m_T\equiv \sqrt{M_{T_{4c}}^2+p_T^2}$. 
To estimate the uncertainties arising from higher-order QCD corrections, we slide the factorization scale in the range $m_T/2 \le \mu_F \le 2 m_T$.
We take the maximum deviation from the center value as the symmetric scale uncertainty.

The numerical predictions for the $p_T$ spectra of various $S$-wave $T_{4c}$ states 
are plotted in Fig.~\ref{fig:pt_dist} and Fig.~\ref{fig:model:compare}. 
In Fig.~\ref{fig:pt_dist}, we compare the $p_T$ distributions of different $T_{4c}$ states, taking both potential models as inputs. 
The lower insets show the ratios of the $\sigma(1^{+-})$ and $\sigma(2^{++})$ states to $\sigma(0^{++})$. 
We observe that,  while the difference between $C$-even states is  minor, the $C$-odd $1^{+-}$ state exhibits much more suppressed cross sections with respect to the $C$-even tetraquarks. 
This observation corroborates the asymptotic $p_T$ scaling behaviors shown in \ref{eq:SDC:large:pT}, where the 
differential cross section for the $1^{+-}$ tetraquark is suppressed by a factor of $1/p_T^{-2}$ with respect to those for the 
$C$-even tetraquarks.

In Fig.~\ref{fig:model:compare}, we compare the predictions made from two phenomenological models. 
Note that two models in general render similar results, except in the case of the $0^{++}$ tetraquark. 
It is the interfering term $2 F_{3,6} \expval{O_{3,6}}$ in (\ref{eq:factorized:cross:section}) that is responsible for the drastic difference. 
As shown in Table~\ref{tab:LDME}, the values of $\expval{O_{3,6}}$ in the two models even take different signs.  Therefore, 
the interfering term is constructive in Model I and destructive in Model II. As a result, the $0^{++}$ state is suppressed in Model II. 
The impact of the interference is demonstrated in Fig.~\ref{fig:pt_comp_dist}, where we single out the individual contributions from different color configuration in 
\ref{eq:factorized:cross:section}. In Model I, the $\triplet$ channel dominates in magnitude, while the other two channels still pose significant positive contributions. 
It is not the case in Model II: the interfering term is negative and the absolute values of all three color channels are roughly the same.

\begin{widetext}
  \begin{minipage}{\linewidth}
    \begin{figure}[H]
      \centering
      \includegraphics[width=0.49\linewidth]{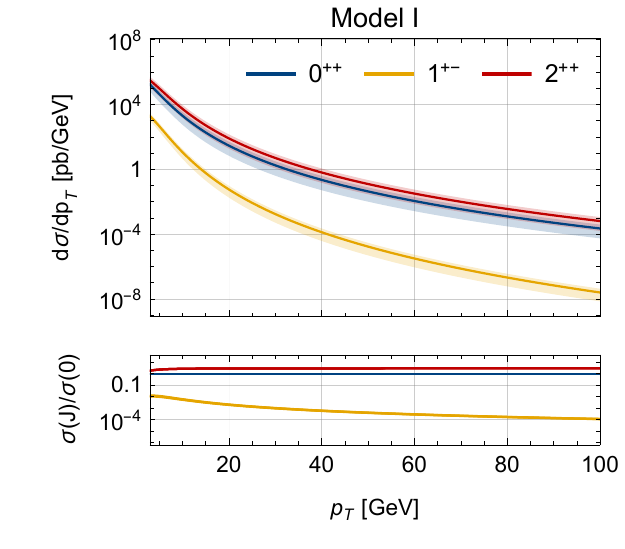}
      \includegraphics[width=0.49\linewidth]{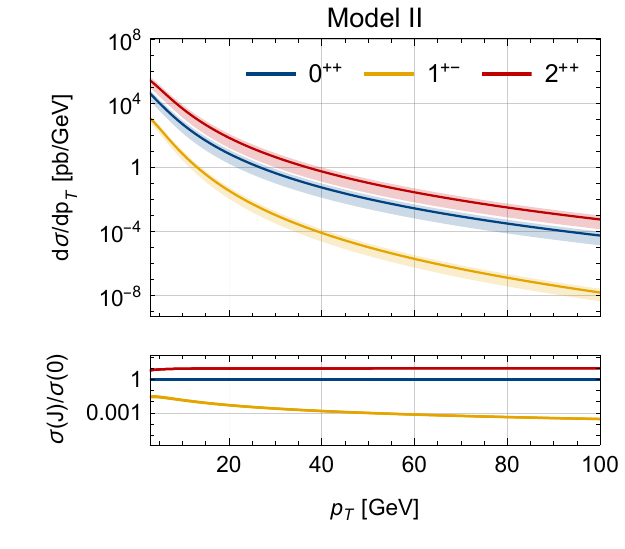}
    \caption{The $p_T$ spectra of the $S$-wave $T_{4c}$ at \texttt{LHC} with $\sqrt{s}=13\ \mathrm{TeV}$ predicted from two potential models.  
    The left panel represents the predictions made from Model I, while the right panel represents the predictions made from Model II. 
    The blue, yellow and red curves represent the differential cross sections for the $0^{++}$, $1^{+-}$ and $2^{++}$ tetraqaurks, 
    respectively. The lower insets show the ratios of $\sigma(1^{+-})$ and $\sigma(2^{++})$ to $\sigma(0^{++})$. }
      \label{fig:pt_dist}
    \end{figure}
  \end{minipage}
\end{widetext}

\begin{widetext}
  \begin{minipage}{\linewidth}
    \begin{figure}[H]
      \centering
      \includegraphics[width=1\textwidth]{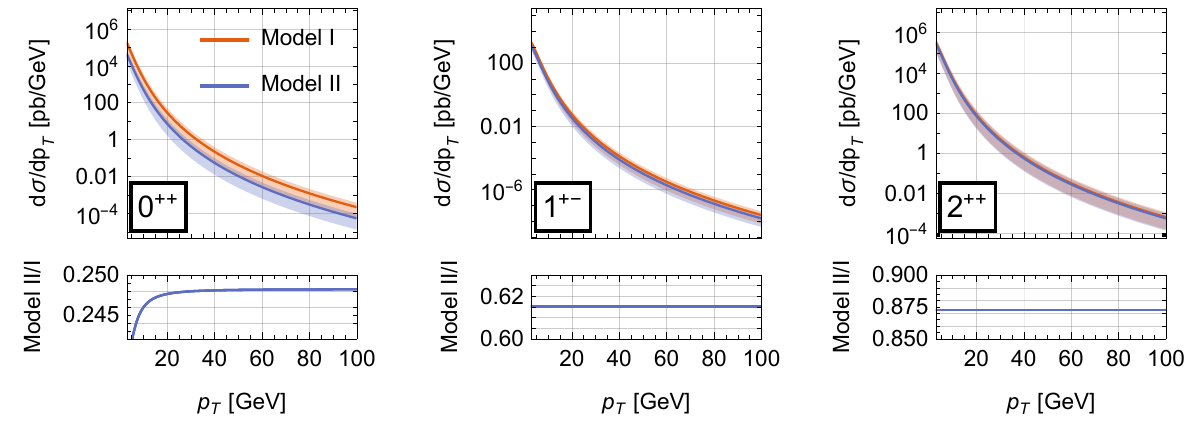}
      \caption{Comparison of the $p_T$ distributions of the $S$-wave $T_{4c}$ between two phenomenological potential models. 
      The left, central and right panels represent the differential cross sections for the $0^{++}$, $1^{+-}$ and $2^{++}$ tetraquarks, respectively. 
      The orange (blue) curves represent the predictions made from Model I (II). The lower insets show the ratios of the predicted production rates in
      Model II to those in Model I.}
      \label{fig:model:compare}
    \end{figure}
  \end{minipage}
\end{widetext}

\begin{widetext}
  \begin{minipage}{\linewidth}
    \begin{figure}[H]
      \centering
      \includegraphics[width=0.49\linewidth]{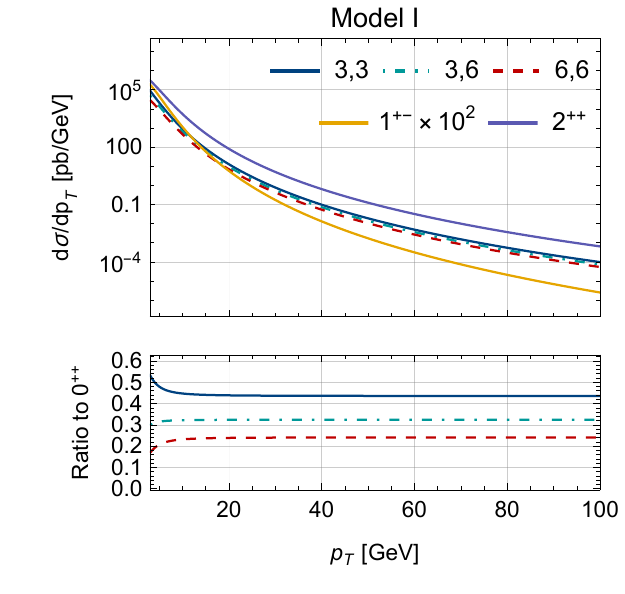}
      \includegraphics[width=0.49\linewidth]{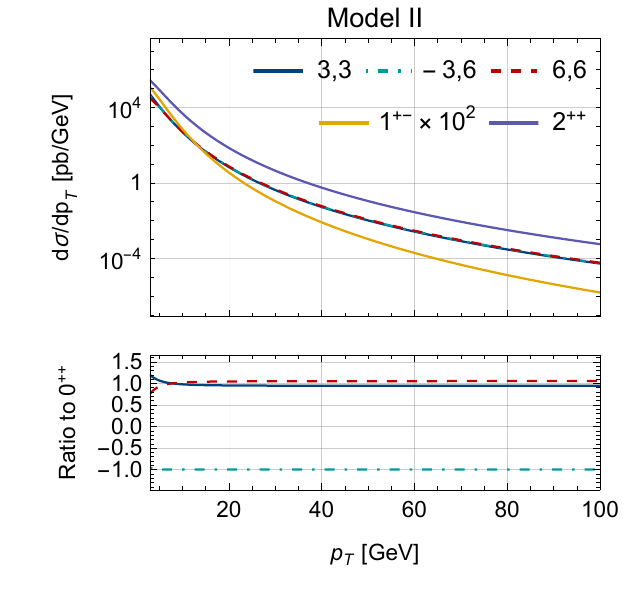}
    \caption{Comparison of contributions from different color configurations in \ref{eq:factorized:cross:section}. The left panel is from Model I, 
    and the right panel is for Model II. The blue solid, green dash-dotted and red dashed curves stand for the contributions from the pure color-triplet, interference and pure color-sextet contributions 
    in $\sigma(0^{++})$, respectively. An additional minus sign is added to the interfering term in Model II to make it positive. 
    The lower insets show the ratio of the individual contributions to the full cross section of the $0^{++}$ tetraquark. 
    We also present the $p_T$ distributions of the $1^{+-}$ and $2^{++}$ states for comparison. }
      \label{fig:pt_comp_dist}
    \end{figure}
  \end{minipage}
\end{widetext}

\begin{widetext}
  \begin{minipage}{\linewidth}
    \begin{figure}[H]
      \centering
      \includegraphics[width=0.49\textwidth]{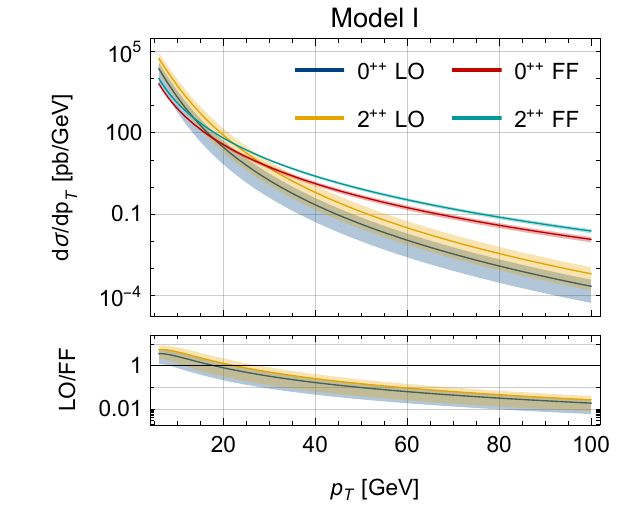}
      \includegraphics[width=0.49\textwidth]{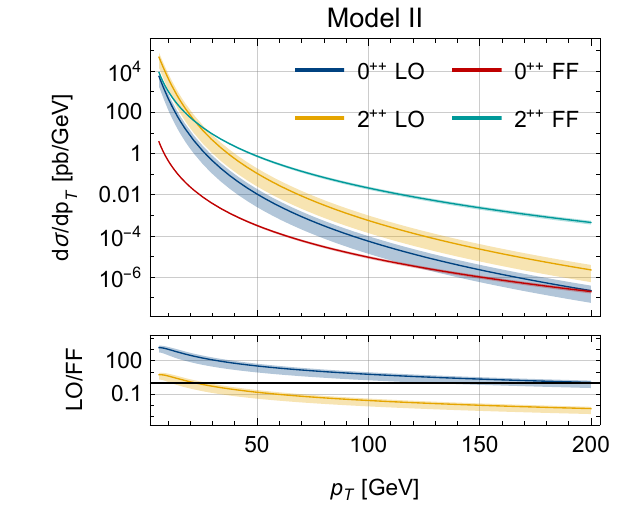}
      \caption{Comparison of the $p_T$ distributions of the $T_{4c}$ between this work and from the fragmentation mechanism~\cite{Feng:2020riv}. 
      The left panel is for Model I, and the right panel is from Model II. The blue and yellow curves represent the 
      leading-order NRQCD predictions for $\sigma(0^{++})$ and $\sigma(2^{++})$, while the red and green curves represent the fragmentation contributions to $\sigma(0^{++})$ and $\sigma(2^{++})$. 
      The lower insets show the ratios of the leading-order NRQCD predictions to the fragmentation predictions. }
      \label{fig:pt_dist:compare}
    \end{figure}
  \end{minipage}
\end{widetext}


In Fig.~\ref{fig:pt_dist:compare}  we also compare our results with the predictions made by the fragmentation mechanism~\cite{Feng:2020riv}. 
As indicated in \eqref{eq:SDC:large:pT}, the fragmentation contribution is enhanced by a factor of $p_T^{2}$ relative to our LO NRQCD predictions at large $p_T$. 
This expectation is confirmed in Fig.~\ref{fig:pt_dist:compare}, where the fragmentation contributions start to overshoot the leading-order results when $p_T\sim 20\ \mathrm{GeV}$.

\begin{table}[H]
    \centering
    \begin{tabular}{ccccc}
      \hline \hline
                & \multicolumn{2}{c}{Model I} & \multicolumn{2}{c}{Model II}                                                        \\
      \cline{2-5}
                & $\sigma\,[\mathrm{nb}]$     & $N_{\mathrm{events}}/10^9$   & $\sigma\,[\mathrm{nb}]$ & $N_{\mathrm{events}}/10^9$ \\
      \hline
      {$0^{++}$} & $37\pm26$      & $110\pm80$  & $9\pm6$       & $27\pm19$     \\
      {$1^{+-}$} & $0.28\pm 0.16$ & $0.8\pm0.5$ & $0.17\pm0.10$ & $0.52\pm0.29$ \\
      {$2^{++}$} & $93\pm65$      & $280\pm200$ & $81\pm57$     & $240\pm170$   \\
      \hline \hline
    \end{tabular}
    \caption{The integrated production rates for various $S$-wave $T_{4c}$ states ($6\,\mathrm{\mathrm{GeV}}\leq p_{T}\leq100\,\mathrm{\mathrm{GeV}}$) and the estimated event yields.}
    \label{tab:total:cross:section}
\end{table}

Finally in Table~\ref{tab:total:cross:section} we present the predicted integrated cross sections for $S$-wave $T_{4c}$ at 
$\sqrt{s}=13\ \mathrm{TeV}$ with a cut $p_T\geq 6\ \mathrm{GeV}$.  
Assuming an integrated luminosity of $3000\ \mathrm{fb}^{-1}$, we also estimate the yields of $T_{4c}$ events at \texttt{LHC}. 
The yields are roughly two orders of magnitude greater than the event yields from fragmentation contributions. 
This is simply due to the fact that the bulk of the $T_{4c}$ cross sections
reside in the low-$p_T$ region.

We note that the $T_{4c}$ cross sections predicted in this work are several orders of magnitude larger than 
the predictions made in Ref.~\cite{Zhu:2020xni}, which range from 10 to  $100\ \mathrm{fb}$.  
Moreover, the ratio of $\sigma(2^{++})$ to $\sigma(0^{++})$ is predicted to be around $2-10$ in this work, is much smaller than that given in Ref.~\cite{Zhu:2020xni} (about $260$), yet relatively closer
to the estimation given in Ref.~\cite{Ma:2020kwb} (about $1-2$).

It is also interesting to compare our predictions for the $1^{+-}$ $T_{4c}$ events with the measured double $J/\psi$ production rate at \texttt{ATLAS}, which is presumed to arise from
double parton scattering~\cite{1410.8822}. 
The predicted cross sections of $T_{4c}$ is almost $50$ times larger than that for double $J/\psi$ production, which is about $5\ \mathrm{pb}$ at $7\ \mathrm{TeV}$

\paragraph{\color{blue}Summary \label{sec:summary}}

In this paper, we predict the $p_T$ spectrum of various $S$-wave fully-charmed tetraquark states within the NRQCD factorization framework, at the lowest order in $\alpha_s$ and velocity. 
The LDMEs are estimated from two phenomenological potential models, with the aid of vacuum saturation approximation. 
The yield of the $1^{+-}$ fully-charmed tetraquark is significantly lower than that for the $0^{++}$ and $2^++$ tetraquarks.
Two potential models render quite different predictions for the differential production rates for the $0^{++}$ tetraquark, signalling the important role played by the interference
between the $\triplet$ and $\sextet$ color channels.
Both models predict that a tremendous number of $T_{4c}$ events would be produced at \texttt{LHC}. It is interesting to await the future experiments measurements 
to confront our predictions.

\begin{acknowledgments}
We are grateful to Ming-Sheng Liu and Qi-Fang L\"u for providing us with the values of tetraquark wave functions at the origin
from their potential models.
The work of F.~F. is supported by the NNSFC Grant No. 12275353, No. 11875318.
The work of Y.-S.~H. is supported by the DOE grants DE-FG02-91ER40684 and DE-AC02-06CH11357.
The work of Y.~J. and J.-Y.~Z. is supported in part by the NNSFC Grants No.~11925506, No.~12070131001 (CRC110 by DFG and NSFC).
The work of W.-L. S. is supported by the NNSFC Grant No.~11975187.
The work of D.-S. Y. is supported by the NNSFC Grant No.~12235008.
\end{acknowledgments}


\end{document}